\documentclass[12pt]{article}
\usepackage{axodraw,bbold}

\catcode`@=12
\begin{document}
\thispagestyle{empty}
\begin{flushright} 
UCRHEP-T441\\ 
October 2007\
\end{flushright}
\vspace{0.5in}
\begin{center}
{\LARGE	\bf SU(5) Completion of the Dark Scalar\\ Doublet
Model of Radiative Neutrino Mass\\}
\vspace{1.5in}
{\bf Ernest Ma\\}
\vspace{0.2in}
{\sl Department of Physics and Astronomy, University of California,\\}
\vspace{0.1in}
{\sl Riverside, California 92521, USA\\}
\vspace{1.5in}
\end{center}

\begin{abstract}\
Adding a second scalar doublet $(\eta^+,\eta^0)$ and three neutral singlet 
fermions $N_{1,2,3}$ to the Standard Model of particle interactions with a 
new $Z_2$ symmetry, it has been shown that $\eta^0_R$ or $\eta^0_I$ is a 
good dark-matter candidate and seesaw neutrino masses are generated 
radiatively. A minimal extension of this new idea is proposed to allow for 
its SU(5) completion.  Supersymmetric unification is then possible, and 
leptoquarks of a special kind are predicted at the TeV scale.
\end{abstract}

\newpage
\baselineskip 24pt

A new idea has recently been proposed that without dark matter, neutrinos 
would be massless.  This is minimally implemented \cite{m06-1,m06-2,m06-3} 
by the addition of a second scalar doublet $(\eta^+,\eta^0)$ and three 
neutral singlet Majorana fermions $N_{1,2,3}$ to the Standard Model (SM) 
of particle interactions, together with a new $Z_2$ discrete symmetry 
\cite{dm78}, under which $(\eta^+,\eta^0)$ and $N_{1,2,3}$ are odd, and all 
SM particles are even.  
Using the allowed term $(\lambda_5/2)(\Phi^\dagger \eta)^2 + H.c.$, where 
$\Phi=(\phi^+,\phi^0)$ is the SM Higgs doublet, seesaw neutrino masses are 
generated in one loop, as shown in Fig.~1.

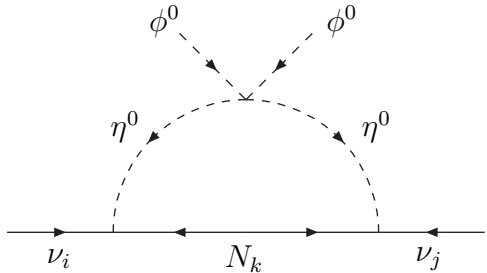
\begin{figure}[htb]
\begin{center}
\begin{picture}(360,120)(0,0)
\ArrowLine(90,10)(130,10)
\ArrowLine(180,10)(130,10)
\ArrowLine(180,10)(230,10)
\ArrowLine(270,10)(230,10)
\DashArrowLine(155,85)(180,60)3
\DashArrowLine(205,85)(180,60)3
\DashArrowArc(180,10)(50,90,180)3
\DashArrowArcn(180,10)(50,90,0)3

\Text(110,0)[]{$\nu_i$}
\Text(250,0)[]{$\nu_j$}
\Text(180,0)[]{$N_k$}
\Text(135,50)[]{$\eta^0$}
\Text(230,50)[]{$\eta^0$}
\Text(150,90)[]{$\phi^{0}$}
\Text(217,90)[]{$\phi^{0}$}

\end{picture}
\end{center}
\caption{One-loop generation of neutrino mass.}
\end{figure}

\noindent At the same time, $\eta^0_R$ and $\eta^0_I$ are split in mass, and 
either is a good dark-matter candidate \cite{bhr06,lnot07,glbe07} with mass 
between 45 and 75 GeV, with reasonable prognosis \cite{bhr06,cmr07} for 
detection at the Large Hadron Collider (LHC). Variants of this basic idea 
have also been discussed \cite{knt03,cs04,kms06,ks06,hkmr07,ss07,bm07,m07}.

In this note, the SU(5) completion of this simplest model is proposed, 
leading to its possible supersymmetric unification. (It has been shown 
\cite{ms07} that the string-inspired $E_6/U(1)_N$ model may also be used, but 
gauge-coupling unification in this case is not as straightforward 
\cite{kmn07}.)  As a consequence, there are two kinds of dark matter, and 
leptoquarks are predicted which always decay into one or both, and may well 
be observable at the LHC.

Under SU(5), the SM quarks and leptons are organized into
\begin{eqnarray}
\underline{5}^* &=& d^c~(3^*,1,1/3) + (\nu,e)~(1,2,-1/2),\\ 
\underline{10} &=& (u,d)~(3,2,1/6) + u^c~(3^*,1,-2/3) + e^c~(1,1,1),
\end{eqnarray}
where their $SU(3)_C \times SU(2)_L \times U(1)_Y$ decompositions are 
also indicated.  If the SM is supersymmetrized and two Higgs superfields 
transforming as
\begin{equation}
\Phi_{1,2} \sim (1,2,\mp 1/2)
\end{equation}
are added, then it is well-known \cite{bs04} that the three SM gauge 
couplings unify at an energy scale of order $10^{16}$ GeV.  This may be 
taken to be an indication of the validity of SU(5) unification.  Note 
in particular that the unification of gauge couplings is insensitive 
to the addition or subtraction of complete SU(5) multiplets, such as in 
split supersymmetry \cite{split}.  In one-loop 
order, each complete multiplet contributes equally to the slopes of 
$\alpha_i^{-1} = 4 \pi/g_i^2$ as a function of energy scale. Hence the 
three gauge couplings $g_i$ still converge at around $10^{16}$ GeV; the 
only change is their numerical value.

In the dark scalar doublet model \cite{m06-1} of seesaw radiative neutrino 
mass, the obvious thing to do is to consider $(\eta^+,\eta^0)$ as part of 
a \underline{5} representation of SU(5), together with its conjugate 
\underline{5}$^*$, i.e.
\begin{eqnarray}
\underline{5} &=& h~(3,1,-1/3) + (\eta_2^+,\eta_2^0)~(1,2,1/2),\\ 
\underline{5}^* &=& h^c~(3^*,1,1/3) + (\eta_1^0,\eta_1^-)~(1,2,-1/2),
\end{eqnarray} 
both of which are of course odd under the new $Z_2$.  In that case, 
gauge-coupling unification is again possible, provided that $m_h$ and 
$m_\eta$ are comparable within an order of magnitude.  For 
$\eta^0_{1,2}$ to be considered as components of dark matter, 
the energy scale of $m_h$ is then likely to be TeV or less. 
Conventionally, the existence of $h(h^c)$ in the \underline{5}
(\underline{5}$^*$) representation of SU(5) is considered dangerous 
because it would mediate rapid proton decay.  However, the new $Z_2$ 
symmetry used here for dark matter and radiative neutrino mass also 
serves the purpose of conserving baryon number and preventing proton 
decay.

As studied already in Ref.~\cite{m06-3}, the $\lambda_5$ term needed 
for Fig.~1 is not available in a supersymmetric context.  Hence one 
additional singlet superfield $\chi$ is needed, as shown in Table 1.

\begin{table}[htb]
\caption{Particle content of proposed model.}
\begin{center}
\begin{tabular}{|c|c|c|}
\hline 
Superfield & $Z_2$ & $Z'_2$ \\ 
\hline
$d^c, (\nu,e)$ & $-$ & + \\ 
$(u,d), u^c, e^c$ & $-$ & + \\ 
$(\phi^0_1,\phi^-_1), (\phi^+_2,\phi^0_2)$ & + & + \\ 
\hline
$N$ & $-$ & $-$ \\ 
$h^c, (\eta^0_1,\eta^-_1)$ & + & $-$ \\ 
$h, (\eta^+_2,\eta^0_2)$ & + & $-$ \\ 
$\chi$ & + & $-$ \\ 
\hline
\end{tabular}
\end{center}
\end{table}

The imposition of $Z_2$ amounts to the usual $R$ parity of the Minimal 
Supersymmetric Standard Model (MSSM).  The additional exactly conserved 
$Z'_2$ forbids the coupling $(\nu \phi_2^0 - e \phi_2^+)N$ but allow 
\cite{m06-3}
\begin{equation}
f_{ij} (\nu_i \eta_2^0 - e_i \eta_2^+) N_j +\lambda_1 \Phi_1 \eta_2 \chi
+ \lambda_2 \Phi_2 \eta_1 \chi + \mu_\phi \Phi_1 \Phi_2 + \mu_\eta \eta_1 
\eta_2 + {1 \over 2} \mu_\chi \chi \chi + {1 \over 2} M_{ij} N_i N_j,
\end{equation}
thereby generating radiative neutrino masses in one loop, as shown in Fig.~2.

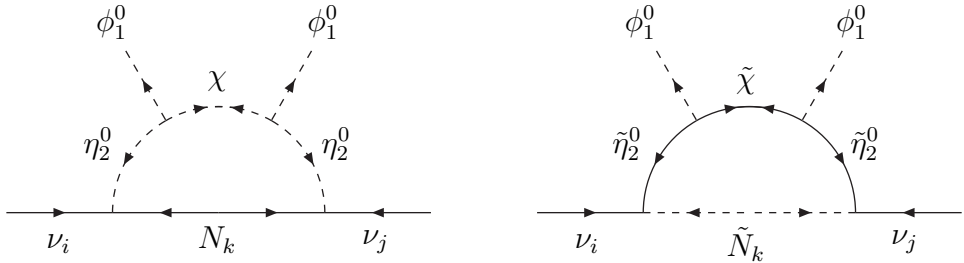
\begin{figure}[htb]
\begin{center}\begin{picture}(500,100)(10,45)
\ArrowLine(70,50)(110,50)
\ArrowLine(150,50)(190,50)
\ArrowLine(150,50)(110,50)
\ArrowLine(230,50)(190,50)
\Text(90,35)[b]{$\nu_i$}
\Text(210,35)[b]{$\nu_j$}
\Text(150,35)[b]{$N_k$}
\Text(150,100)[]{$\chi$}
\Text(105,70)[b]{$\eta^0_2$}
\Text(195,70)[b]{$\eta^0_2$}
\Text(110,116)[b]{$\phi^0_1$}
\Text(190,116)[b]{$\phi^0_1$}
\DashArrowLine(130,85)(115,111){3}
\DashArrowLine(170,85)(185,111){3}
\DashArrowArc(150,50)(40,120,180){3}
\DashArrowArc(150,50)(40,60,100){3}
\DashArrowArcn(150,50)(40,60,0){3}
\DashArrowArcn(150,50)(40,120,80){3}

\ArrowLine(270,50)(310,50)
\DashArrowLine(352,50)(390,50){3}
\DashArrowLine(348,50)(310,50){3}
\ArrowLine(430,50)(390,50)
\Text(290,35)[b]{$\nu_i$}
\Text(410,35)[b]{$\nu_j$}
\Text(350,32)[b]{$\tilde N_k$}
\Text(350,100)[]{$\tilde \chi$}
\Text(305,70)[b]{$\tilde \eta^0_2$}
\Text(395,70)[b]{$\tilde \eta^0_2$}
\Text(310,116)[b]{$\phi^0_1$}
\Text(390,116)[b]{$\phi^0_1$}
\DashArrowLine(330,85)(315,111){3}
\DashArrowLine(370,85)(385,111){3}
\ArrowArc(350,50)(40,120,180)
\ArrowArc(350,50)(40,60,100)
\ArrowArcn(350,50)(40,60,0)
\ArrowArcn(350,50)(40,120,80)

\end{picture}
\end{center}
\caption[]{One-loop radiative contributions to neutrino mass.}
\end{figure}

Because of the two separately conserved discrete symmetries, there are now 
at least two absolutely stable particles: the lightest particle with $R=-1$ 
as in the MSSM, and the lightest particle which is odd under $Z'_2$. 
Consider in particular the three lightest particles with $(R,Z'_2) = (-,+)$, 
$(+,-)$, and $(-,-)$ respectively. If one is heavier than the other two 
combined, then the latter are the two components of dark matter.  If not, 
then all three contribute. 

The new prediction of the proposed supersymmetric SU(5) completion of the 
dark scalar doublet model is of course the leptoquark $h(h^c)$ and its 
associated scalar partners $\tilde h$ and $\tilde h^c$.  Because they are 
odd under $Z'_2$, they must decay into one or both of the dark-matter 
candidates with $(R,Z'_2) = (+,-)$ or $(-,-)$.  If kinematically allowed, 
the $(\pm,-)$ particle may also decay into $(\mp,-)$ and $(-,+)$.
Specifically, the only allowed trilinear coupling involving $h$ or $h^c$ 
is $h d^c N$.  Assuming that $N$ is much heavier than $h$, then from Eq.~(6) 
it is clear that $h$ always decays into a quark and a lepton plus a particle 
which is odd under $Z'_2$.

Since $m_h$ should be at the TeV scale or below (from the argument that the 
gauge couplings should be unified), strong production of $h \bar{h}$ at 
the LHC is expected.  Note however that $h$ is different from the usual 
leptoquark which conserves both additive baryon number $B$ and lepton number 
$L$. Here $h$ has $B = 1/3$, but only odd $(-)^L$.  This means that it can 
decay into either $d e^- \eta_2^+$ or $d e^+ \eta_2^-$.  Thus $h \bar{h}$ 
production will result in same-sign dileptons plus quark jets plus missing 
energy, which is a possible unique signature of this exotic particle.

In conclusion, by adding the new exotic leptoquark superfields $h$ and $h^c$ 
to a supersymmetric version of the dark scalar doublet model of radiative 
seesaw neutrino mass, the SU(5) completion of the model is accomplished, 
allowing the gauge couplings to be unified as in the MSSM.  The pair 
production of these leptoquarks will result in same-sign dilepton events 
with missing energy, which may be observable at the LHC.

This work was supported in part by the U.~S.~Department of Energy under Grant 
No.~DE-FG03-94ER40837.

\baselineskip 16pt

\bibliographystyle{unsrt}

\end{document}